\documentclass[12pt]{iopart}

\usepackage{graphicx}  
\graphicspath{{fig/}}
\begin{document}

\title[ Hyperon suppression in hadron-quark mixed phase]{
 Hyperon suppression in hadron-quark mixed phase}

\author{Toshiki Maruyama$^{1*}$, Satoshi Chiba$^1$, Hans-Josef Schulze$^2$ and Toshitaka Tatsumi$^3$}

\address{$^1$ Advanced Science Research Center, Japan Atomic Energy Agency, Tokai, Ibaraki, 319-1195, Japan}
\address{$^2$ INFN Sezione di Catania, Catania, Sicily, I-95123, Italy}
\address{$^3$ Department of Physics, Kyoto University, Kyoto, 606-8502, Japan}

\address{$*$ E-mail: \tt maruyama.toshiki@jaea.go.jp}

%
%
%

\begin{abstract}

\noindent
We investigate the property of the hadron-quark mixed phase 
using the Brueckner-Hartree-Fock model for hadron (hyperon) phase
and the MIT bag model for quark phase.
To satisfy the Gibbs conditions, charge density as well as baryon number density
becomes non-uniform in the mixed phase, accompanying phase separation.
We clarify the roles of the surface tension and the charge screening effect. 
We show that the screened Coulomb interaction
tends to make the geometrical structure of the mixed phase less stable,
and the resultant EOS becomes similar to the one given by
the Maxwell construction.
The composition of the mixed phase, however, is very different
from that of the Maxwell construction;
in particular, hyperons are completely 
suppressed in the mixed phase, because hadron phase is positively charged. 
This is a novel mechanism of hyperon suppression in compact stars.

\medskip
\noindent
(Some figures in this article are in color only in the electric version)

\end{abstract}

\vspace{2pc}


\section{Introduction}

\def\sgm{\Sigma^-}

\def\tsurf{\sigma}
\def\vc{V_{\rm C}}
\def\rv{\bi{r}}

\def\ms{M_{\odot}}
\def\bc{B=100\;\rm MeV/fm^{-3}}

It is well known that 
at several times the normal nuclear density $\rho_0$,
hyperons emerge in matter and lead to a strong softening of the equation of state (EOS).
Consequently the maximum neutron star mass is reduced to the one much lower   
than currently observed values of $\sim 1.4M_\odot$.
For example the microscopic Brueckner-Hartree-Fock approach
gives much lower masses.

On the other hand, the hadron-quark deconfinement transition is believed 
to occur in hot and/or dense matter.
Then one may expect the maximum mass increases to the Chandrasekhar limit once 
the deconfinement transition occurs in hyperon matter \cite{hypns}.
The deconfinement transition from the hadron to quark phase may 
be of first order.
It brings about a thermodynamic instability 
of uniform matter to have phase separation.
In other words, matter should have the nonuniform mixed phase (MP) around
the critical density. 
Since hadron and/or quark matter consists of many kinds of particles, 
the Gibbs conditions must be properly taken into account. 
The usual Maxwell construction (MC) can be no more applied in this case. 
Due to the interplay of the Coulomb interaction 
and the surface tension between two phases, the MP can have exotic
shapes called ``pasta'' structures \cite{rav} (as a review, see Ref.\ \cite{mar}).
With increase of density, the stable shape of pasta structures 
may change from droplet to rod, slab, tube, and to bubble. 
The name ``pasta'' comes from rod and slab structures 
figuratively spoken as ``spaghetti'' and ``lasagna''.


Generally, the appearance of the MP in matter results in a softening of the EOS.
The bulk Gibbs calculation (BG) of the MP, 
without the effects of the Coulomb interaction and surface tension, 
leads to an appearance of the MP in a broad density region \cite{gle92}. 
If one takes into account the geometrical structures of the MP, however,
the EOS deviates from that of the BG.
It approaches to the one given by 
the MC \cite{mar}.

In this report
we explore the EOS and the structure of the MP during the
hyperon-quark transition, 
properly taking account of the Gibbs conditions together with the pasta structures.
Then we demonstrate a novel phenomenon, suppression of hyperons in the MP. 

\section{Numerical Calculation}

 The numerical procedure to determine the EOS and the
 geometrical structure of the MP is 
 explained in detail in Ref.~\cite{maruhyp}.
We employ the Wigner-Seitz cell approximation in which
the whole space is divided into equivalent 
cells with a given geometrical symmetry,
sphere for three dimensional (3D) case, cylinder for 2D, and slab for 1D.
In each cell the quark and hadron phases are spatially separated 
by a sharp boundary.
The energy density of the MP is then written as
%
\begin{equation}
 \!\!\epsilon = {1\over {V_W}} \left[ 
 \int_{V_H} \!\!d{\rv}\, \epsilon_H({\rv})+
 \!\int_{V_Q} \!\!d{\rv}\, \epsilon_Q({\rv})+
 \!\int_{V_W} \!\!d{\rv} \left( \!\epsilon_e({\rv}) \!+ \!{(\nabla \vc({\rv}))^2\over 8\pi e^2} \right)
 \!+ \!\tsurf S \right]\!,
\end{equation}
where the volume of the Wigner-Seitz cell $V_W$ is the sum of 
those of hadron and quark phases $V_H$ and $V_Q$, and 
$S$ the quark-hadron interface area.
The surface energy is taken into account with a surface-tension parameter $\tsurf$.
The quantities $\epsilon_H$, $\epsilon_Q$ and $\epsilon_e$ 
are energy densities of hadrons, quarks and electrons, respectively, 
which are 
functions of
local densities 
of $n,p,\Lambda,\Sigma^-,u,d,s,e$. 
For a given density $\rho_B$, the optimum configuration of the cell 
(uniform hadron, quark droplet, rod, slab, tube, bubble, or uniform quark),
the cell size $R_W$, the lump size $R$,
and the density profile of each component
are searched for to give the minimum energy density.
%

To calculate $\epsilon_H$ 
we use the Thomas-Fermi approximation for the kinetic energy density.
The interaction-energy density is 
calculated by the nonrelativistic BHF approach \cite{hypns}
based on the microscopic
NN and NY potentials.
With these potentials, the various $G$ matrices are evaluated 
by solving numerically the Bethe-Goldstone equation, which can be written in 
operator form as 
\begin{equation}
 G_{ab}[W] = V_{ab} + \sum_c \sum_{p,p'} V_{ac} \big|pp'\big\rangle 
 {Q_c \over W - E_c +i\epsilon} 
  \big\langle pp'\big| G_{cb}[W] \:, 
\label{e:g}
\end{equation}
where the indices $a,b,c$ indicate pairs of baryons
and the Pauli operator $Q$ and energy $E$ 
determine the propagation of intermediate baryon pairs.
The pair energy in a given channel $c=(ij); i,j=n,p,\Lambda,\Sigma$ is
\begin{equation}
 E_{(ij)} = T_{i}(k_{i}) + T_{j}(k_{j})
 + U_{i}(k_{i}) + U_{j}(k_{j})
\label{e:e}
\end{equation}
with 
$T_i(k) = m_i + {k^2\!/2m_i}$.
The various single-particle potentials are given 
self-consistently from the $G$ matrices as,
\begin{equation}
 U_i(k) = 
 \sum_{j=n,p,\Lambda,\Sigma} \sum_{k'<k_F^{(j)}} 
  {\rm Re} \big\langle k k' \big| G_{(ij)(ij)}\big[E_{(ij)}(k,k')\big] 
  \big| k k' \big\rangle  \:.
\label{e:un}
\end{equation}
Once the different single-particle potentials are known,
the total nonrelativistic hadronic energy density, $\epsilon_H$,  
can be evaluated:
\begin{eqnarray}
 \epsilon_H \!&=& \!\!\!\!
 \sum_{i=n,p,\Lambda,\Sigma} 
 \sum_{k<k_F^{(i)}}
 \left[ T_i(k) + {1\over2} U_i(k) \right] \:,
\label{e:eps}
\end{eqnarray}
and $\epsilon_H$ is thus represented as a function of particle number densities
$\rho_i (i=n,p,\Lambda,\Sigma)$. 
%
The parameter set used in this calculation
reproduces the scattering phase shifts and nuclear saturation property.

For the quark phase,
we employ the MIT bag model with 
massless $u$ and $d$ quarks and massive $s$ quark with $m_s= 150$ MeV.
The energy density $\epsilon_Q$ consists of the kinetic term by the Thomas-Fermi approximation,
the leading-order one-gluon-exchange term  
and the bag constant $B$ as
\begin{eqnarray}
 \epsilon_Q &=& B + \sum_f \epsilon_f \:,
\\
 \epsilon_f(\rho_f) &=& {3m_f^4 \over 8\pi^2} \bigg[ 
 { x_f\left(2x_f^2+1\right)\sqrt{1 + x_f^2}} - {\rm asinh}\,x_f \bigg] 
\nonumber \\
 && - \alpha_s{m_f^4\over \pi^3} \bigg[ 
 x_f^4 - {3\over2}\Big( x_f \sqrt{1 + x_f^2} - {\rm asinh}\,x_f \Big)^2 
 \bigg] \:,
\end{eqnarray}
where $m_f$ is the $f$ current quark mass,
$x_f = k_F^{(f)}\!/m_f$,
and the number density of $f$ quarks $\rho_f = {k_F^{(f)}}^3\!\!/\pi^2$.

Demanding that the quark EOS crosses the hadronic EOS at
a reasonable density, 
we choose $B=100$ $\rm MeV/fm^3$ and $\alpha_s=0$.

The surface tension of the hadron-quark interface is poorly known, 
but some theoretical estimates based on the MIT bag model 
for strangelets 
and lattice gauge simulations at finite temperature 
suggest a range of $\tsurf \approx 10$--$100\;\rm MeV\!/fm^2$.
We employ $\tsurf=40\;\rm MeV\!/fm^2$ 
in the present study.

\section{Hadron-Quark Mixed Phase}

\begin{figure*}[b]
\includegraphics[width=0.99\textwidth]{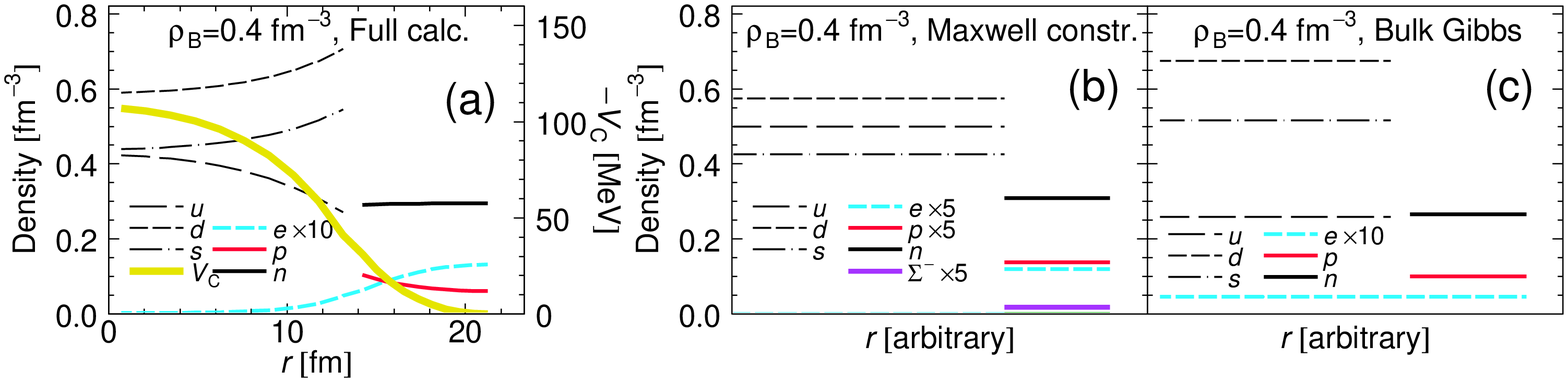}
\vspace{-4mm}
\caption{
(a) Density profiles 
and Coulomb potential $\vc$ 
within a 3D (quark droplet) Wigner-Seitz cell
of the MP at $\rho_B=0.4$ fm$^{-3}$.
(b) Same as (a) for MC case.
The radius $r$ is in arbitrary unit. 
(c) The case of BG calculation.
}
\label{figProf}
\end{figure*}

Figure~\ref{figProf} (a) illustrates an example of 
the density profile in a 3D cell for $\rho_B=0.4$ fm$^{-3}$.
One can see the non-uniform density distribution of each particle species
together with the finite Coulomb potential. 
Charged particle
distributions are rearranged by the Coulomb potential.
For example,
the quark phase is negatively charged, so that 
$d$ and $s$ quarks are repelled to the boundary
of the negatively charged quark phase, 
while $u$ quarks gather at the center.
The protons in the hadron phase are attracted by the negatively charged 
quark droplet, while the electrons mostly exist in the hadron phase.
This density rearrangement of the charged particles 
causes the screening of the Coulomb interaction between two phases.

In panels (b) and (c), depicted are the cases of MC and BG for comparison.
MC assumes the local charge neutrality,
while the BG does not.
One can see that the local charge neutrality in the full calculation 
lies between two cases.

\section{Effects of the Coulomb Screening and the Surface Tension}

\begin{figure}
\includegraphics[width=0.4\textwidth]{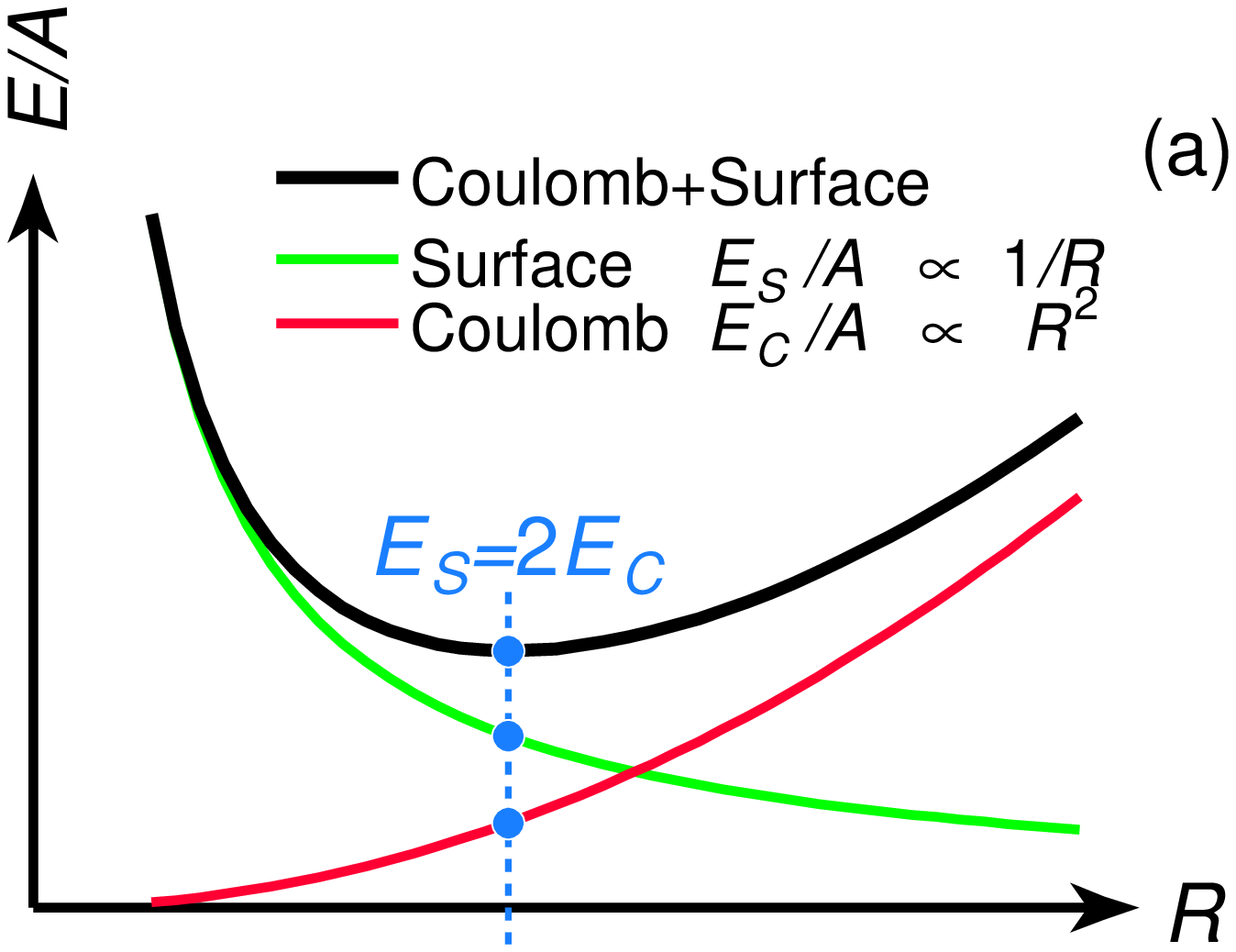}
\includegraphics[width=0.44\textwidth]{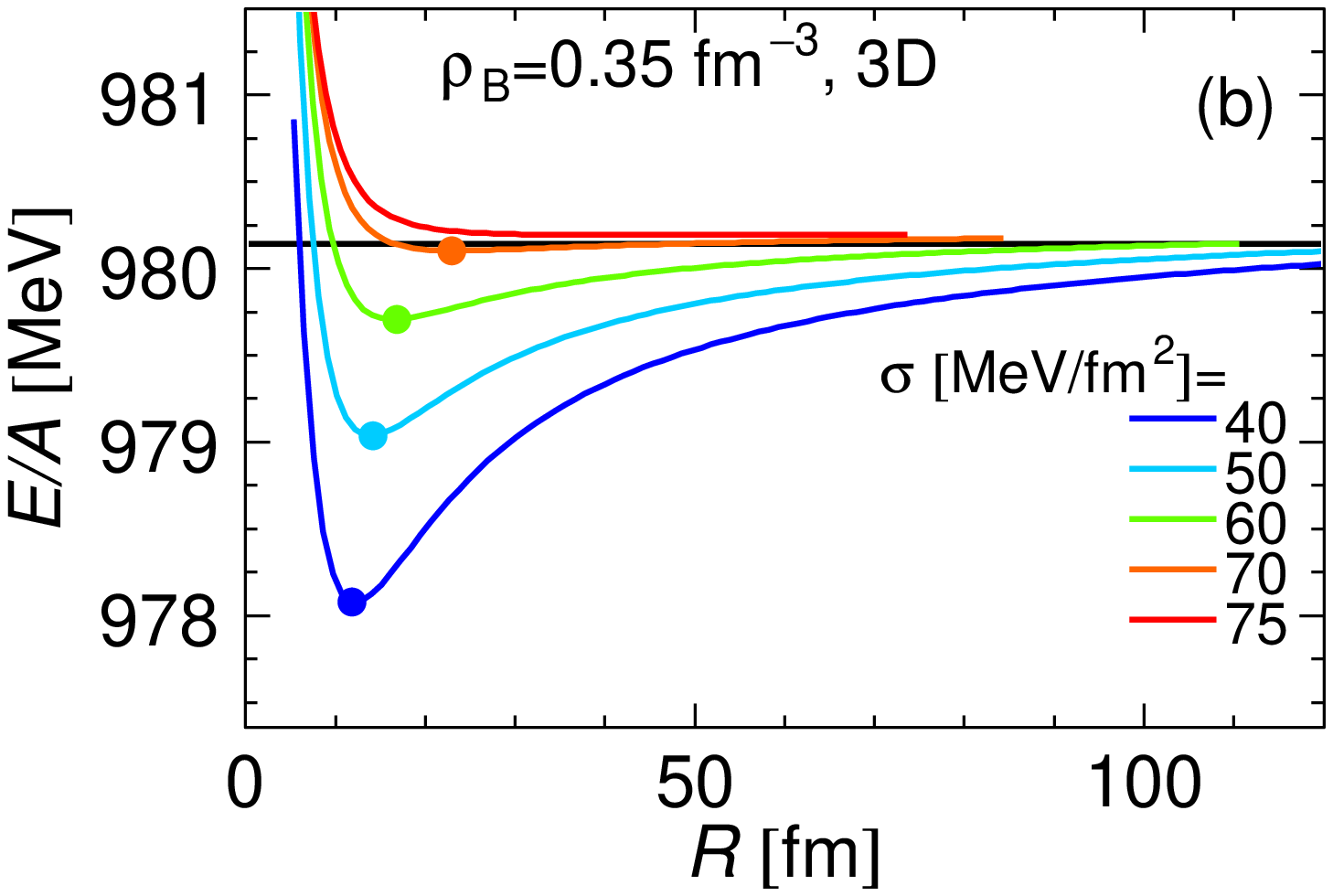}
\caption{
(a) 
Energy per particle from the Coulomb and surface contributions 
for a 3D droplet structure.
The volume fraction is fixed with a variation of the structure size.
(b) Total energy per baryon of droplet structure in the full calculation. 
}
\label{figBalance}
\end{figure}

Here let us discuss the effects of the Coulomb screening 
and the surface tension.
The volume fraction $V_{i}/V_W$ ($i=Q,H$) is determined by a bulk
calculation without any surface tension or the Coulomb interaction.   
Then the size of the 
structure is determined by the
balance of the Coulomb repulsion and the surface tension,
as schematically explained in Fig.~\ref{figBalance}~(a).
Taking three dimensional case for example,
the Coulomb energy per particle $E_C/A$ is roughly proportional
to the second power of the structure size $R$,
while the surface energy per particle $E_S/A$ is 
to the inverse of $R$.
Then the sum of these two energies has a minimum
at a certain value of $R$ with $E_S=2E_C$ \cite{rav}.

As the $E_C/A$ curve gets lower, the minimum point moves to 
the right hand side, i.e., the structure size becomes large.
In the same way, if the surface tension is stronger (weaker), 
the structure size becomes larger (smaller).

When the screening of the Coulomb interaction is incorporated, 
the situation should be changed: the $R^2$ dependence
breaks and $E_C/A$ decreases to zero as $R\rightarrow \infty$. 
Therefore the minimum point of $(E_C+E_S)/A$ disappears, 
if both the screening effect
and the surface tension become strong enough.
%
%
In Fig.~\ref{figBalance}~(b) we demonstrate an example, 
where the energy per particle is depicted by changing 
the strength of the surface tension. 
One can see that there is no minimum 
for $\tsurf>70$MeV.

\section{Maxwell Construction and the Bulk Gibbs Calculation}

\begin{figure}[b]
\includegraphics[width=0.51\textwidth]{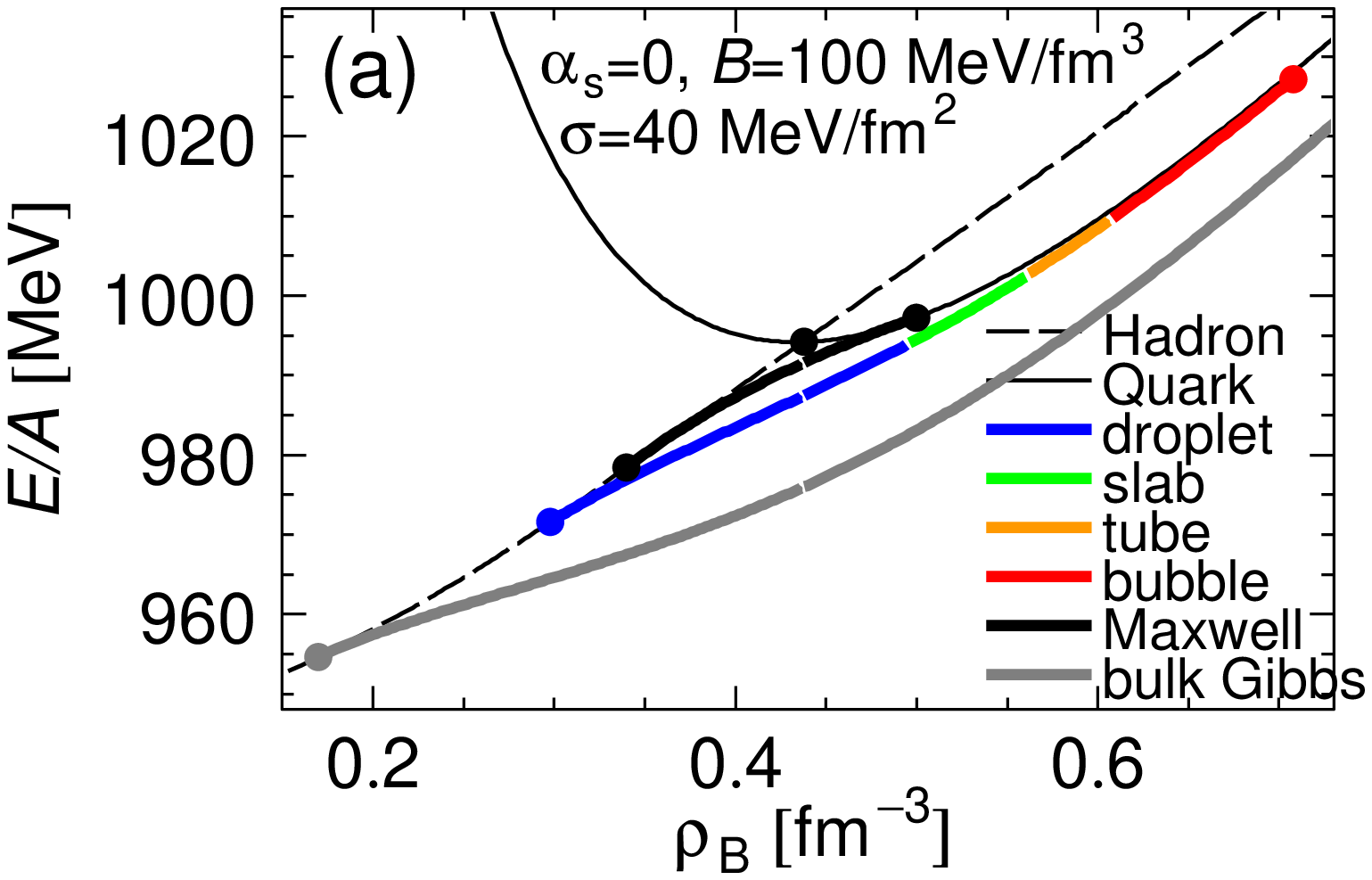}
\includegraphics[width=0.48\textwidth]{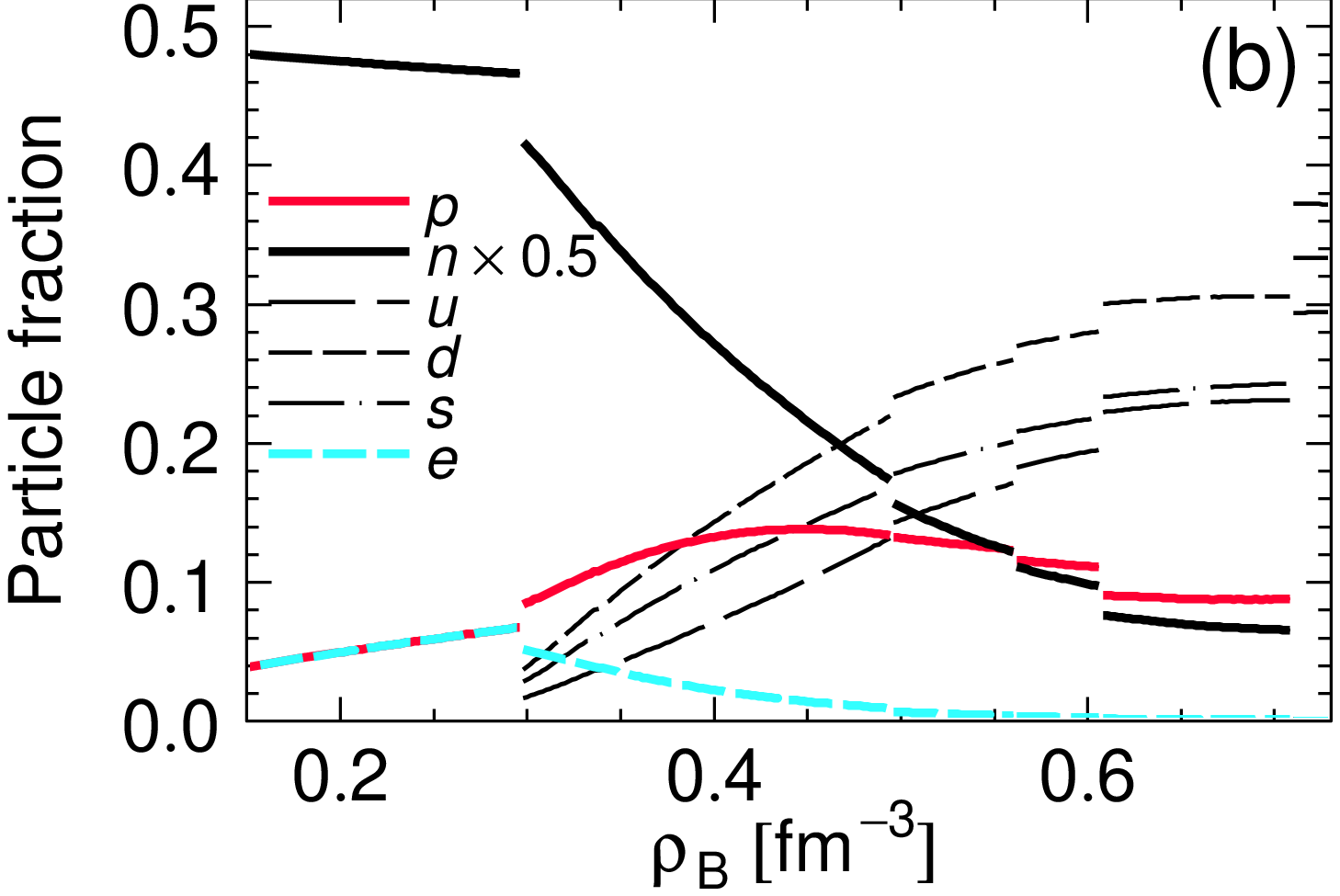}
\caption{
(a) 
The energy per baryon 
of the MP (thick curves)
in comparison with pure hadron and quark phases (thin curves),
MC (black) and BG (gray) calculations.
(b) 
Particle fractions 
in the MP 
by the full calculation.
}
\label{figEOS}
\end{figure}

Figure~\ref{figEOS}~(a) compares the resulting EOS 
with that of the pure hadron and quark phases.
The thick black curve indicates the case of the MC,
while the colored lines indicate the MP
with various geometries
starting at $\rho_B=0.326$ fm$^{-3}$ 
with a quark droplet structure
and terminating at $\rho_B=0.666$ fm$^{-3}$ 
with a quark bubble structure.
Note that the charge screening effect always tends to 
make matter locally charge-neutral and to reduce the Coulomb energy. 
Combined with the surface tension, it 
makes the non-uniform structures mechanically less stable
and limits the density region of the MP \cite{mar,maruhyp}.
Figure~\ref{figEOS}~(b) shows the particle fraction by the full calculation.
One can see that there appears no hyperon in matter.
We shall discuss this point later.

Next let us consider the dependence on the surface tension.
If the surface tension is strong, the structure size 
becomes large.
In the larger scale, the Coulomb screening effect becomes more prominent
and the local charge neutrality will be approximately achieved.
Consequently the energy of the MP is 
close to that of the MC. 
On the other hand, if the surface tension is weak, the structure size
becomes small.
Then the Coulomb interaction becomes ineffective. 
Therefore the aspects of the MP becomes close to
that of the BG.

If we use $\tsurf\approx 60\ \rm MeV/fm^2$, the EOS of the MP will 
coincide with the MC (thick black curve).
If we use small value of $\tsurf$, it approaches to the BG (thick gray curve).
Our surface tension parameter $\tsurf=40\ \rm MeV/fm^2$
is strong enough for the MP to be close to the MC case.

\section{Structure of Hybrid Stars}

\begin{figure}[b]
\includegraphics[width=0.50\textwidth]{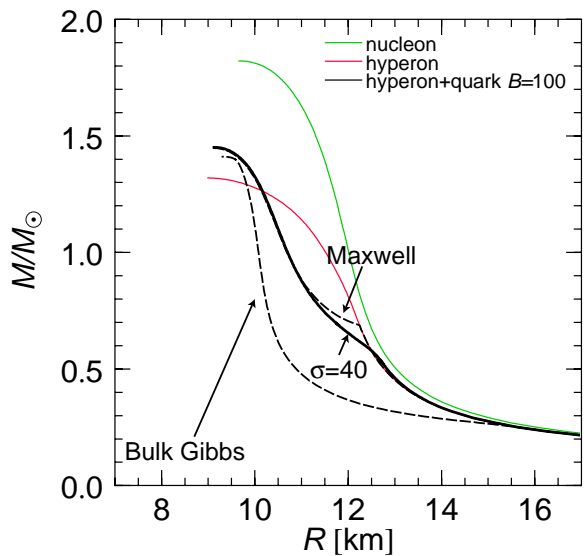}
\vspace{-5mm}
\caption{
Neutron star mass-radius relations for different EOS.
For the hybrid stars,
the dashed lines indicate the MC (upper curve)
or BG calculation (lower curve) 
and the solid line the MP of the full calculation.
}
\label{f:mr}
\end{figure}

Knowing the EOS comprising hadronic, mixed, and quark phase
in the form $P(\epsilon)$,
the equilibrium configurations of static NS are obtained
in the standard way
by solving the Tolman-Oppenheimer-Volkoff (TOV)  equation for 
the pressure $P(r)$ and the enclosed mass $m(r)$,
\begin{eqnarray}
  {dP\over dr} &=& -{ G m \epsilon \over r^2 } \,
  {  \left( 1 + {P / \epsilon} \right) 
  \left( 1 + {4\pi r^3 P / m} \right) 
  \over
  1 - {2G m/ r} } \:,\qquad
\\
  {dm \over dr} &=& 4 \pi r^2 \epsilon \:,
\end{eqnarray}
with $G$ being the gravitational constant. 
Starting with a central mass density $\epsilon(r=0) \equiv \epsilon_c$,  
one integrates out until the surface density equals the one of iron.
This gives the stellar radius $R$ and its gravitational mass $M=m(R)$.

Our EOS gives results similar to those given by the MC.
The maximum mass of a hybrid star is around $1.5\,M_\odot$, larger than
that of the purely hadronic (hyperonic) star, $\approx1.3\,M_\odot$.
Hence we may conclude that a hybrid star is still consistent with the 
canonical NS mass of $1.4\,M_\odot$, 
while the masses of purely hyperonic stars lie below it.

\section{Suppression of Hyperons}


The structure and the composition of the MP, however, are very different from 
those of the MC.
Though a relevant hyperon ($\Sigma^-$) fraction is finite in the MC case,
it is completely suppressed up to very high density in the full calculation
(see Fig.~\ref{figEOS} (b)).
%
%
%
\begin{figure}
\includegraphics[width=0.73\textwidth]{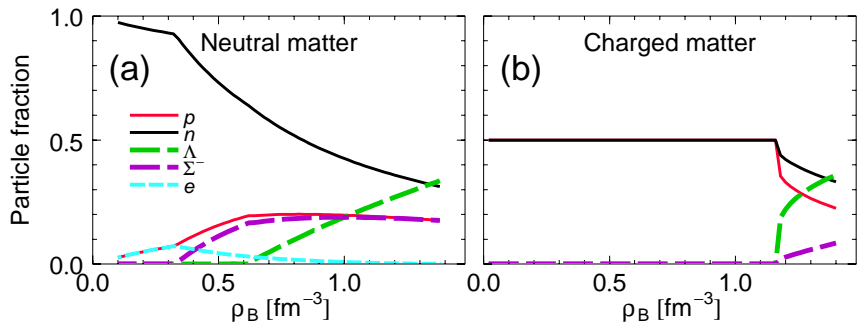}
\vspace{-3mm}
\caption{
(a) Particle fractions of neutral matter with electrons. 
(b) The same quantity for charged matter without electrons,
the low-density part of which corresponds to symmetric nuclear matter.
}
\label{figRatioUnif}
\end{figure}
\begin{figure}[t]
\includegraphics[width=0.49\textwidth]{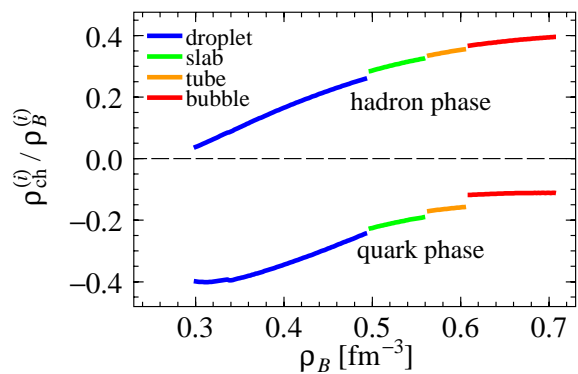}
\vspace{-3mm}
\caption{
Charge number density per baryon number density in each phase 
$\rho_{\rm ch}^{(i)}/\rho_B^{(i)}$ ($i=Q, H$)
as a function of baryon number density.
Neutral hadron matter, for instance. has $\rho_{\rm ch}^{(H)}/\rho_B^{(H)}=0$,
and symmetric nuclear matter 0.5.
}
\label{figcharge}
\end{figure}
The suppression of hyperon mixture in the MP
is due to the fact that the hadron phase is positively charged.
As shown in Fig.\ \ref{figRatioUnif}, 
hyperons ($\Sigma^-$) appear in charge-neutral hadronic matter 
at low density ($0.34\;\rm fm^{-3}$)
to reduce the Fermi energies of electron and neutron.
In the absence of the charge-neutrality condition, on the other hand, 
symmetric nuclear matter will be realized at lower density
and hyperons will be mixed  
at higher density (above $1.15\;\rm fm^{-3}$) 
due to the large hyperon masses.
Although the Coulomb screening effect diminishes the local charge density,
the MP has positively charged hadron phase and negatively charged quark phase.
Thus, the mixture of hyperons is suppressed in the MP where the hadron phase is positively charged.

In Fig.\ \ref{figcharge} we plot charge number density per baryon number density in hadron and quark phases. 
It is clear that the hadron/quark phase is positively/negatively charged.
At $\rho_B\approx 0.3$ $\rm fm^{-3}$ where quark phase has a
very small volume fraction, negative charge of the quark phase is large.
This is because it should compensate
the positive charge of the hadron phase with a large volume fraction.
On the other hand, at high densities where hadron phase has small volume fraction,
the hadron phase is positively charged and the value approaches to that of 
symmetric nuclear matter.
This high charge of hadron phase at high density leads to the suppression of hyperons.

\section{Summary}

In this article we have studied the properties of the MP 
in the quark deconfinement transition in hyperonic matter, and their
influence on compact stars.

The hadron-quark MP was consistently treated with the
basic thermodynamical requirement due to the Gibbs conditions. 
We have seen that the resultant EOS
is close to the one given by the MC. 
This is because the finite-size effects, i.e.\ the strong surface tension  
and the Coulomb screening,  
enlarge the structure size and promote the local charge neutrality.
They also tend to diminish the density region 
of the MP through the mechanical instability.
The masses and radii of compact stars given by our EOS are  
similar to those given by the MC. 
The maximum mass of a hybrid star is around $1.5\,M_\odot$, larger than
that of the purely hadronic star,  $\approx1.3\,M_\odot$. 

Although the EOS of matter and the resultant bulk properties of
compact stars are close to those in the case of the MC, 
the ingredients of the MP are found to be different.
Hyperons are completely suppressed in the MP.
This is a novel feature of the hadron-quark MP
and 
should have important consequences for the elementary processes 
inside compact stars. 
For example, coherent scattering of neutrinos off lumps in the
MP may enhance the neutrino opacity \cite{red}. 
Also, the absence of hyperons prevents a fast cooling mechanism by way of 
the hyperon Urca processes \cite{pet,ttt,tny}. 
These results directly modify the thermal evolution of compact stars.


\end{document}